\documentstyle[epsf,11pt]{article}

\newcommand{\grs}{$\gamma$-rays }
\newcommand{\gr}{$\gamma$-ray }

\newcommand{\ale}{\alpha_{\rm e} }

\newcommand{\alr}{\alpha_{\rm r} }
\newcommand{\alinj}{\alpha_{\rm inj} }

\begin {document}

\begin{centering}

{\Large Study of the synchrotron and inverse Compton 

radiations of relativistic jets in GRS 1915+105}

\vspace{5mm}

{\large A.M.Atoyan$^{1,2}$ and F.A.Aharonian$^{1}$}

\vspace{3mm}

\end{centering}

{\small
\hspace{33mm} (1)~~Max-Planck-Institut f\"ur Kernphysik, Heidelberg, Germany

\hspace{33mm} (2)~~Yerevan Physics Institute, Yerevan, Armenia
}

\vspace{5mm}

\begin{centering}

{\bf Abstract}

\end{centering}

{\small
\vspace{1mm}
\noindent
We show that comparison of the detailed calculations of the 
synchrotron radiation of  expanding magnetized clouds (plasmoids) 
with the temporal
and spectral evolution of radio flares observed from the 
recently discovered galactic ``microquasar'' GRS 1915+105
imposes strong constraints on the model parameters of relativistic 
plasmoids, in particular, on the absolute values  and time-dependence of 
the magnetic field, speed of expansion of radio clouds, the 
rates of injection of relativistic electrons into and their energy-dependent 
escape from the clouds, etc. 
The data of radio monitoring of the pair of ejecta not 
only enable unambiguous determination of the aspect angle and the speeds of 
propagation of both components, 
but also contain an important  information 
about the energy source for acceleration of the electrons, in particular, may 
distinguish between the scenarios of bow-shock powered and magnetized 
relativistic wind powered plasmoids. 
Within the framework of the synchrotron-self-Compton model 
we also calculate the fluxes of synchrotron hard X-rays and 
GeV/TeV inverse Compton 
$\gamma$-rays,
and discuss constraints on the parameters of GRS~1915+105 which could be  
provided by positive detection or flux upper limits of these energetic photons.
 }

\vspace{5mm}

\noindent
{\bf 1~~Introduction}
\vspace{3mm}

\noindent
The galactic black hole (BH) candidate source GRS 1915+105, discovered in 
1992 as a hard X-ray transient (Castro-Tirado et al. 1994), 
attracts an extreme  
interest since the detection of superluminal radiojets in this source 
by Mirabel \& Rodriguez (1994;
hereafter MR94). 
Together  with the second galactic 
superluminal jet source  GRO J1655-40 ( Tingay et al. 1995; Hjellming \& 
Rupen, 1995), they constitute a new class of objects called   
{\it microquasars} (MR94), representing a scaled 
down analogs of QSOs (or generally AGNs) in two principal
aspects: both microquasars and quasars are most probably powered 
by accretion onto BHs, although of essentially different mass
scales, and both populations  produce relativistic jets. 
Being, however,
much closer to us than AGNs, the microquasars enable radio monitoring 
of not only approaching but also receding ejecta (importantly, in short 
timescales), which makes them unique cosmic laboratories
for study of the phenomenon of relativistic jets (MR94). 

Here we summarize the basic results of our recent study 
(Atoyan \& Aharonian, 1997; AA97) 
of the prominent March/April 1994 radio flare of GRS 
1915+105, when
the approaching ({\it south}) and receding ({\it north}) ejecta moving in 
opposite directions with a speed $\beta \simeq 0.92$  at an aspect angle 
$\theta \simeq 70^{\circ}$ have been resolved, and 
superluminal speed 
$1.25\,c$ (for $d=12.5\,\rm kpc$) has been discovered (MR94).   

\vspace{4mm}
\noindent
{\bf 2~~General scenario}

\vspace{2mm}
\noindent

\noindent
 Comprehensive discussion of radio flares observed from GRS 1915+105, 
which most probably are connected with 
ejection of radio emitting material (i.e. clouds,
containing relativistic electrons and 
magnetic fields frozen in  the co-moving plasma), is given in 
MR94, Rodriguez et al. (1995), Foster et al. (1996). Here we would emphasize
two important features of time evolution of the flares: 

\noindent
({\bf a}) a rapid, 
less than a day, rise time of the fluxes, $S_\nu \propto \nu^{-\alr}$, 
revealing transition from optically thick
($\alr \leq 0$) to
thin ($\alr > 0$) emission, which is a typical signature of an expanding 
radio source; 

\noindent
({\bf b}) the spectral index $\alr\sim 0.5$ at the stages of flare
maximum later on steepens to $\alr \sim 1$. 

\vspace{2mm}
The study of these  features results in conclusive 
information on the intrinsic parameters of the radio source.
Indeed, modeling the ejecta (in its rest frame) as a spherical source of a 
radius $R$ and magnetic field $B$, with a power-law distribution of 
relativistic electrons  $N(\gamma)\propto \gamma^{-2}$ extending to $\gamma
\geq 10^4$, we can express 
$R$ and $B$ in terms of $S_{\ast} \equiv S_{10\,\rm GHz}/500\,\rm mJy$
and cloud's opacity  
$\tau_{\nu}=R\times \kappa_{\nu}$ ($\kappa_{\nu}$ is the synchrotron 
absorption coefficient)
at a frequency $\nu$ as (AA97): 
\begin{equation}
B \simeq 0.15 \, \delta^{7/17} \eta^{-4/17} S_{\ast}^{-2/17}
\tau_{\nu}^{6/17}\nu_{\rm GHz}^{18/17} \; \rm G \, ,
\end{equation}
\begin{equation}
R = \, 4.6\times 10^{14} \, \delta^{-28/17}\eta^{-1/17} 
S_{\ast}^{8/17}
\tau_{\nu}^{-7/17} \nu_{\rm GHz}^{-21/17} \,\rm cm\; ,
\end{equation}
where $\nu_{\rm GHz} \equiv \nu/1\,\rm GHz$, $\delta = \sqrt{1-\beta^2 }/
(1-\beta \cos \theta) \simeq 0.57$ is the Doppler factor,
and $\eta= w_{\rm e} /w_{\rm B}$ is the ratio of the electron to magnetic 
field energy densities. For the outburst 
of 19 March 1994, the fluxes measured at instant $t_0
\simeq 4.8\,\rm days$  after ejection 
(the time of VLA observations on March 24.6), 
corresponded to $S_\ast \simeq 1$. Therefore, as it follows from Eq.(2), 
for optical transparency $\tau_{\nu} \leq 1$ apparent down to 
$\nu \simeq 1.4\,\rm GHz$, one needs $R\geq 8\times 10^{14}\,\rm cm$. 
To reach that size during $t_{0}^{\prime}=\delta \,t_0 \simeq 2.7\,\rm days$, 
the radio clouds should expand in their rest frame with a speed 
$v_{\rm exp} >0.1\, c\,$. This agrees with
the speed of expansion deduced directly from 
observations, and suggests that the lack of detection of  
red-shifted optical lines might be due to large Doppler-broadening of the
lines  (Mirabel et al. 1997). 

For $\tau_{1.4\,\rm GHz}\leq 1$~~Eq.(1) results in
$B\leq 0.2 \eta^{-4/17}\,\rm G$ at times $t\simeq t_0$.  
For the equipartition ($\eta = 1$) magnetic field 
$B_{\rm eq}\sim 0.2\,\rm G$ the synchrotron cooling time of   
electrons emitting at frequencies $\nu\sim 10\,\rm GHz$
exceeds few years. Therefore the synchrotron
losses cannot be responsible for the observed steepening of 
radio spectra to $\alr \simeq 1$, which requires a steepening of the electron 
spectrum 
$N(\gamma,t)$ to an index $\ale \simeq 3$.  
We suggest a model (AA97) which  attributes the steepening of $N(\gamma,t)$  
to (i) {\it continuous injection} of relativistic electrons 
into the radio cloud, with the rate $Q(\gamma,t)\propto 
\gamma^{-2}q(t)$, 
{\it and} (ii) {\it energy-dependent  escape} 
$\tau_{\rm esc}(\gamma,t)\propto \gamma^{-1}$
of the electrons from the cloud.

\vspace{4mm}
\noindent
{\bf 3~~The March/April radio f\,lare of GRS 1915+105 }
\vspace{3mm}

\noindent
For calculations of the nonthermal radiation 
expected during radio flares, we have to find temporal evolution of 
the energy distribution of electrons  
$N(\gamma,t)$ in an expanding magnetized cloud. Generally, the
equation for $N(\gamma,t)$, including the therm for stochastic 
{\it in-situ} acceleration,
represents the partial differential equation of the second order 
(e.g. see Kardashev 1962). 
However, in the cases when the region of particle acceleration
can be distinguished from the region of bulk emission, the acceleration
terms can be effectively replaced by injection of 
accelerated particles into the radiation production region, which results in 
a well known  
differential equation of the first order:  
\begin{equation}
\frac{\partial N}{\partial t}\, = \, \frac{\partial}{\partial \gamma}
[P(\gamma,t) N]
\, -\, \frac{N}{\tau_{\rm esc}(\gamma,t)} \, +\, Q(\gamma,t)  \; . 
\end{equation}

For the parameters characterizing the radio clouds in GRS~1915+105, the energy
losses $P=-({\rm d} \gamma / {\rm d} t)$ of electrons in an expanding 
magnetized medium are mainly due to adiabatic losses, and at very high   
energies also the radiative losses. Assuming the temporal evolution
of magnetic field in the form of $B(t)=B_0 (R/R_0)^{-m}$, where  
$R=R(t)$ and $R_0 \equiv R(t_0)\,$, we have $P = 
v_{\rm exp}\,\gamma/R + p_2 \gamma^2/R^{2m}$ ($p_2=\rm const$). 
Assuming $v_{\rm exp} =\rm const$, we have found (AA97) analytical solution 
to Eq.(3). This solution can be used in numerical calculations for an arbitrary 
time dependence of $v_{\rm exp}(t)$, if approximating the latter by the
mean $\bar{v_{\rm i}}$ in  small time intervals $\Delta t_{\rm i}$. 

Allowing for a (natural) deceleration of cloud's expansion at later 
stages, we suppose $v_{\rm exp} = v_0/(1+t^\prime/t_{\rm exp})^k$, where
$t^\prime$ is in the rest frame of the ejecta. The 
injection rate is taken as $Q(\gamma,t^\prime)\propto 
\gamma^{-\alinj} \exp(-\gamma/\gamma_{\rm c})\,q(t^\prime)\,$, which 
implies that the shape of 
energy spectrum (power law with exponential cutoff) does not change in time. 
The term $q(t^\prime) = (1+t^{\prime}/t_{\rm inj} )^{-p}$ 
defines  the time profile of the electron injection rate. 
For the escape time we suppose  energy
dependence in the form   
$\tau_{\rm esc}(\gamma,t^\prime) \propto R^{-u}\gamma^{-\Delta }$ and take
into account that $\tau_{\rm esc}\geq R/c$.

\begin{figure}
\epsfxsize=16. cm
\epsffile[45 153 562 473]{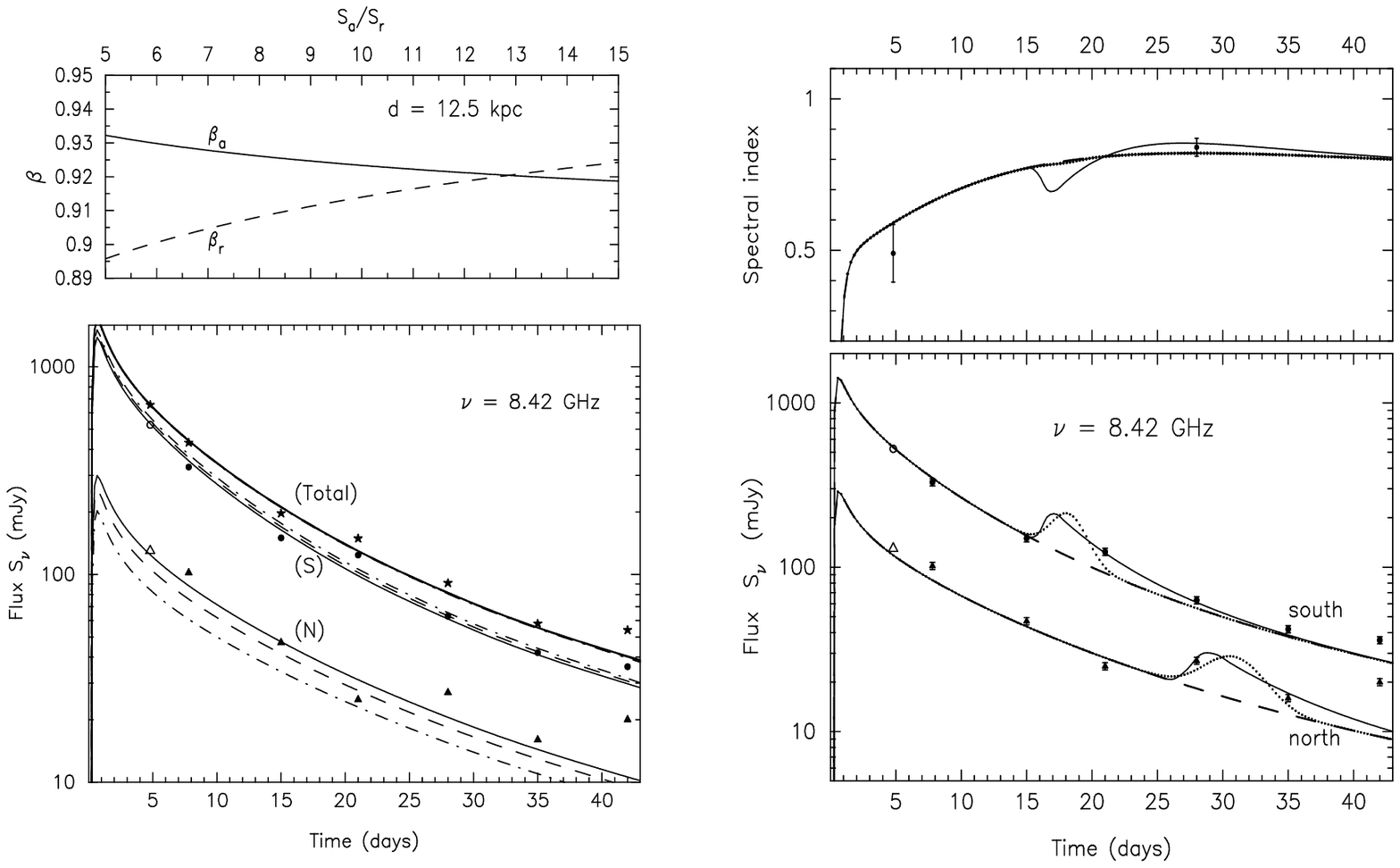}
{{\protect \bf Fig.1 a }~(left):~{\protect \it Top panel} {\small -- 
The speeds of approaching (solid line) and receding (dashed line) ejecta 
calculated from Eq.(4) for different ratios of $S_{\rm a}/S_{\rm r}$. 
In the symmetric case, 
$\beta_{\rm a}=\beta_{\rm r}\,(=0.920)$, the aspect angle 
is $\theta =69.6^\circ$, resulting in
$S_{\rm a}/S_{\rm r}=12.8$;}
{\protect \it Bottom panel} {\small --Evolution of the 
radio fluxes from the 
south (S) and north (N) radio components  expected in the case of 
identical ejecta with $\beta_{\rm a} = \beta_{\rm r}$ 
(dot-dashed lines), and slightly asymmetrical ones, with $S_{\rm a}/S_{\rm b}
=7$ (solid lines) and 9 (dashed lines). Note that the total fluxes coincide.}\\ 
{\protect \bf Fig.1 b}~(right): \small The time evolution of the radio fluxes 
and spectral indexes at $8.42\,\rm GHz$, expected in 
the case of synchronous increase (by factor of 2) of the injection rate of 
relativistic electrons into the south and north radio clouds
during $\Delta t\leq 1\,\rm day$ around {\it equal} 
intrinsic times $t^\prime =
9\,\rm days$ (solid curves). The dashed curves show the spectral evolution  
without an `afterimpulse'. The dotted curves correspond to 
the case when the `afterimpulse' is in the form of an increase of 
magnetic field by $40\,\%$, while the injection of electrons proceeds 
smoothly. To distinguish the flux curves, somewhat 
different shapes of the `magnetic' and `electronic' afterimpulses are supposed.
The {\it principal} difference between these 2 cases is in different time 
evolution of $\alr$.} 
\end{figure}

\vspace{3mm}
\noindent
\underline{\it Model parameters}
\vspace{2mm}

Comparison of the numerical calculations of 
synchrotron radiation of radio clouds with the fluxes detected during 
March/April flare (MR94; Rodriguez et al. 1995)  
prove that initial speed of expansion $v_0\sim (0.1-0.2)\, c$ is needed, 
depending on the magnetic field $B_0$ at instant $t_0
= 4.8\,\rm days$ after ejection. 
For the equipartition magnetic field, $B_{\rm eq}\simeq 0.2\,\rm G$, a 
typical power of injection of relativistic electrons 
$L_{\rm e}\sim (0.5-1) \times 
10^{38}\,\rm erg/s$ is required. However, $L_{\rm e}$ should be much higher
if $B_0 < B_{\rm eq}$, e.g., $L_{\rm e} \geq 10^{39}\,\rm erg/s$
for $B_0\simeq 0.05\,\rm G$.

The observed rate of decline of the flare requires 
a decrease of $B(R)\propto R^{-m}$ 
with $m\simeq 1$. 
The power law index $\alinj \simeq 2$ follows 
from the observation of $\alr\simeq 0.5$ on
March 24,
when the escape losses have not yet significantly modified the electron 
spectrum $N(\gamma,t)$. The observed steepening of the radio spectrum 
at $\sim 10\,\rm GHz$ from $\alr \simeq 0.5$ to $\alr\simeq 0.84$ on April 16 
requires, in our model, an escape of the electrons with 
$\Delta \simeq 1$, and allows for some variation of 
the parameter $u\sim
1.5$. Importantly, the steepening of radio spectra can be 
explained only if the cloud's expansion after
$t_{\rm exp} \sim (1-3) \,\rm days$ gradually decelerates, 
with $k\sim (0.7-1)$, {\it and} 
if about the same time (namely, $t_{\rm inj}\sim (1-2)\,t_{\rm exp}$) the
injection of new particles declines with $p\simeq 2k$.
Otherwise the escape of radio electrons from rapidly expanding cloud would not
be fast enough to steepen $N(\gamma,t)$ as needed. Remarkably, an 
injection rate of this kind can be easily provided, in particular, if one 
would assume that it is proportional to the solid angle of the cloud
as observed from the central source: $q(t^\prime) \rightarrow 
q_{\rm b}(t^\prime)= [R(t^\prime)/v_0 t^\prime]^2$. 
We prefer to call $q_{\rm b}(t^\prime)$ 
as the `beam injection' case, and further on use this type of 
injection in order to reduce the number of free model-parameters.

\vspace{3mm}

\noindent
\underline{\it Brightness ratio as a measure of South/North asymmetry}
\vspace{2mm}

To explain the observed (MR94) brightness ratio of the 
approaching to receding ejecta $S_{\rm a}/S_{\rm r} = 8\pm 1$ (instead of
$S_{\rm a}/S_{\rm r} \geq 12$ expected for {\it identical} 
plasmoids), 
we suppose that the twin ejecta are {\it similar} but not  identical 
(for another possibility see Bodo \& Ghisellini, 1996),
allowing for some asymmetry between the pair of
ejecta in the speeds of propagation ($\beta_{\rm a}\neq \beta_{\rm r}$), 
aspect angles ($\theta_{\rm r}\neq 180^{\circ} -\theta_{\rm a}$) or 
intrinsic luminosities ($S_{\rm a }^{\prime}\neq S_{\rm r }^{\prime}$). 
In particular, the degree of South/North asymmetry needed is smallest
if mainly the speeds 
$\beta_{\rm a}$ and  $\beta_{\rm r}$ would be different. In this case  
for the ratio $S_{\rm a}/S_{\rm r}$ at equal {\it intrinsic times} 
we have 
\begin{equation}
\frac{S_{\rm a}}{S_{\rm b}} = \left( \frac{\delta_{\rm a}}{\delta_{\rm r}}
\right)^{3+\alr} =
\left(\frac{\Gamma_{\rm r}}{\Gamma_{\rm a}}\right)^{3+\alr}
\left(\frac{1+\beta_{\rm r} \cos \theta}{1-\beta_{\rm a} \cos \theta}
\right)^{3+\alr} \;\cdot
\end{equation}
Due to strong dependence of Eq.(4) on the ratio of Lorentz-factors 
$\Gamma_{\rm r}/\Gamma_{\rm a}$, a deviation of the ratio 
$S_{\rm a}/S_{\rm r}$ from the one expected for {\it identical} ejecta 
can be explained by a small difference between $\beta_{\rm a}$
and $\beta_{\rm r}$ (see top panel in Fig.1a). 
For example, solution of Eq.(4)
together with the equations describing the apparent angular separation of the 
ejecta, results in $\beta_{\rm a}=0.928$, $\beta_{\rm r}=0.905$, 
$\theta=70.3^{\circ}$ for $S_{\rm a}/S_{\rm r}=7$.  
In this case the fluxes of both components seem to agree with the ones  
measured by MR94 (see Fig.1a, bottom panel). 
Note that for 
the given parameters of GRS~1915+105, the ratio $S_{\rm a}/S_{\rm r}=7$
at equal {\it intrinsic times} corresponds to  
$S_{\rm a}/S_{\rm r}\simeq 8$ measured at equal {\it angular separations}.

\vspace{3mm}
\noindent
\underline{\it Synchronous afterimpulses far away from the central source ?}
\vspace{3mm}

The reported accuracy
of VLA flux measurements in MR94 is $\sim 5\,\%$. Meanwhile some of the data 
points in Fig.1a, in particular the fluxes on 21-st 
day (April 9) for the south, 
and 28-th day (April 16) for the north components,  exceed the 
calculated fluxes by $\gg 5\,\%$, therefore they still need a better 
explanation. 
We would like to believe that this discrepancy could be connected with 
significant increase of the fluxes between April 4 and 5 ($t=16-17\,\rm days$) 
detected by the Nancay telescope. 
It is seen from Table 1 and Fig.1 in Rodriguez et al (1995), that  
during 24 hours between these 2 days  the fluxes at 
both $1.41\,\rm GHz$ and $3.3 \,\rm GHz$ have 
suddenly {\it increased} by 
$\simeq 30\,\%$. Although VLA was not observing GRS~1915+105 
at that days, the `echo' of that event could be present in the flux detected
by VLA from the approaching component on April 9. Also, one should    
expect significant {\it delay} between the times of  VLA 
observations of that event
from the {\it south} and {\it north} radio clouds, if it was due to powerful 
two-sided   
afterimpulse  from the central source which has reached the two
clouds at equal intrinsic times $t_{\rm a}^{\prime}= t_{\rm r}^{\prime}$. 
In Fig.1b we show the fluxes calculated  for practically the same model 
parameters as in Fig.1a (for $S_{\rm a}/S_{\rm r}= 7$), but assuming that
there was an additional short ($\Delta t^\prime \leq 1\,\rm day$) impulse of 
injection 
of relativistic electrons into {\it both} clouds around intrinsic times
$t^\prime =9\,\rm days$ after ejection event. 
Taking into account different Doppler factors of
counter ejecta, this intrinsic time corresponds to 
{\it apparent} times $t_{\rm a}=16.6\,\rm days$ and $t_{\rm r}=27.5\,\rm days$
for the approaching and receding components, respectively.
It is seen from Fig.1b that agreement with the measured fluxes 
now becomes better. Note that the last two data points 
(on April 30) cannot be explained by another afterimpulse, since they coincide
in time. The excess fluxes on that day are most probably connected with
the second ejection event occured, as noted in MR94, around April 23. 

If such an interpretation of the data is not just an
artefact, but corresponds to reality, implications for the physics of jets
may be very important: it could mean that both
clouds continue to be energized by the central source being even far away from 
it (presumably, through a relativistic wind in the jet region, and perhaps 
also the wind
termination {\it reverse} shock on the {\it back} side of the clouds), 
or otherwise we 
have to rely upon a mere coincidence of equal intrinsic times 
(as well as amplitudes) of additional injection of relativistic particles
from the bow shocks ahead of two counter ejecta. 

\vspace{4mm}
\noindent
{\bf 4~~Predictions for nonthermal high energy radiation}
\vspace{3mm}

\noindent
GRS 1915+105 is a very strong transient X-ray source, with  peak luminosities
exceeding $10^{39}\,\rm erg/s$. This radiation most probably originates
in the thermal plasma near the central compact object.
In addition, nonthermal X-rays of synchrotron origin
could be produced in the relativistic ejecta, if the spectrum of 
accelerated electrons there  
would extend beyond TeV energies as it is the case in blazars 
(e.g. see Urry \&
Padovani 1995). Although from the point of view of required energetics in jets
it is obvious that the X-radiation of the jet generally cannot be responsible
for the bulk of observed X-ray fluxes, the
synchrotron component, however, may become visible at high energies,
say $E_{\rm X} \geq 100\,\rm keV$, where the 
thermal emission is typically suppressed.

In Fig.2 we
present the time evolution of the synchrotron and IC 
radiation fluxes in the range from radio to very high
energy $\gamma$-rays expected in the case of cloud's  
magnetic field $B_0=0.05\,\rm G$, calculated for the injection 
spectrum of electrons with exponential cutoff around $E_{\rm c}=30\,\rm TeV$.
It is seen that during the first several hours after ejection
(or somewhat more, depending on the actual time needed for
formation/acceleration of the expanding ejecta), when the cloud is still opaque
for synchrotron emission at GHz frequencies, the synchrotron fluxes may 
dominate over the extrapolation
of thermal component 
at energies $\geq 100\,\rm keV$. This may result in significant flattening of
overal spectrum at $E_{\rm X}\sim 100\,\rm keV$.
Interestingly,  hard tails of X-ray spectra perhaps are a common
feature of the galactic BH source population, among which is
GRS~1915+105 (Harmon et al. 1994).
Although an existence of such feature in the spectrum of GRS 1915+105 could be
seen only marginally (see e.g. Sazonov et al. 1994), in case of
the second microquasar GRO J1655-40 the X-ray spectra clearly extend 
up to several hundred keV (see Harmon et al. 1995). 

The second signature of acceleration of TeV electrons in GRS~1915+105
could be detection of IC $\gamma$-rays. The calculations in the framework
of synchrotron-self-Compton models, based on the parameters defined from
radio observations, show that during the strong flares we may expect 
detectable fluxes of \grs, first of all at TeV energies, provided that the
magnetic field in the radio clouds does not exceed $0.1\,\rm G$.
In Fig.2 we show the fluxes of IC \grs calculated for parameters of 
the March/April 1994 flare at 3 different times after the outburst.
We can see that the  
$\gamma$-ray fluxes above several hundred GeV 
during first few hours of the flare 
are on the level of $3\times 10^{-11}\,\rm erg/cm^2 s$, which is about of
the 
VHE \gr flux of the Crab Nebula. After few days the flux drops to 
the level of $0.1\,\rm Crab$ which is still detectable by current
Imaging Cherenkov telescopes in Northern hemisphere (CAT, HEGRA, Whipple).
Afterwards the source becomes invisible. 

Assumption of lower magnetic field results in higher (approximately 
$\propto B^{-2}$) \gr fluxes. However,  
as far as a magnetic field
significantly less than $0.05\,\rm G$ would imply injection luminosity
of electrons $L_{\rm e} \geq 10^{40} \,\rm erg/s$, it is difficult
to expect \gr fluxes essentially exceeding the fluxes shown in Fig.2. 
On the other hand, in the case of magnetic field on the level of equipartition
$B\simeq 0.2\,\rm G$ or higher, the 
IC fluxes dramatically decrease. 
Therefore either positive detection or upper limits of VHE \gr fluxes, 
being combined with hard X-ray observations above 100\,keV, could provide
robust constraints on the magnetic field in the ejecta (or the luminosity 
$L_{\rm e}$) and/or 
efficiency of acceleration of electrons beyond TeV energies.

Information about IC \grs can be obtained also in EGRET energy domain,
i.e. at $E\geq 100\,\rm MeV$. However, it should be noted that even during the 
first few hours of the flare, the fluxes in this energy range do not exceed 
$10^{-6}\,\rm ph/cm^2 s$. This implies that with the effective detection area
of the EGRET $\sim 10^3\,\rm cm^2$, only several photons could be detected
during the first day of the outburst.
Unfortunately, since the IC fluxes significantly drop
with expansion of the cloud,  
observations of the flare after that time 
could not
noticeably improve the photon statistics.

\begin{figure}
\vspace{4.8cm}
\includegraphics{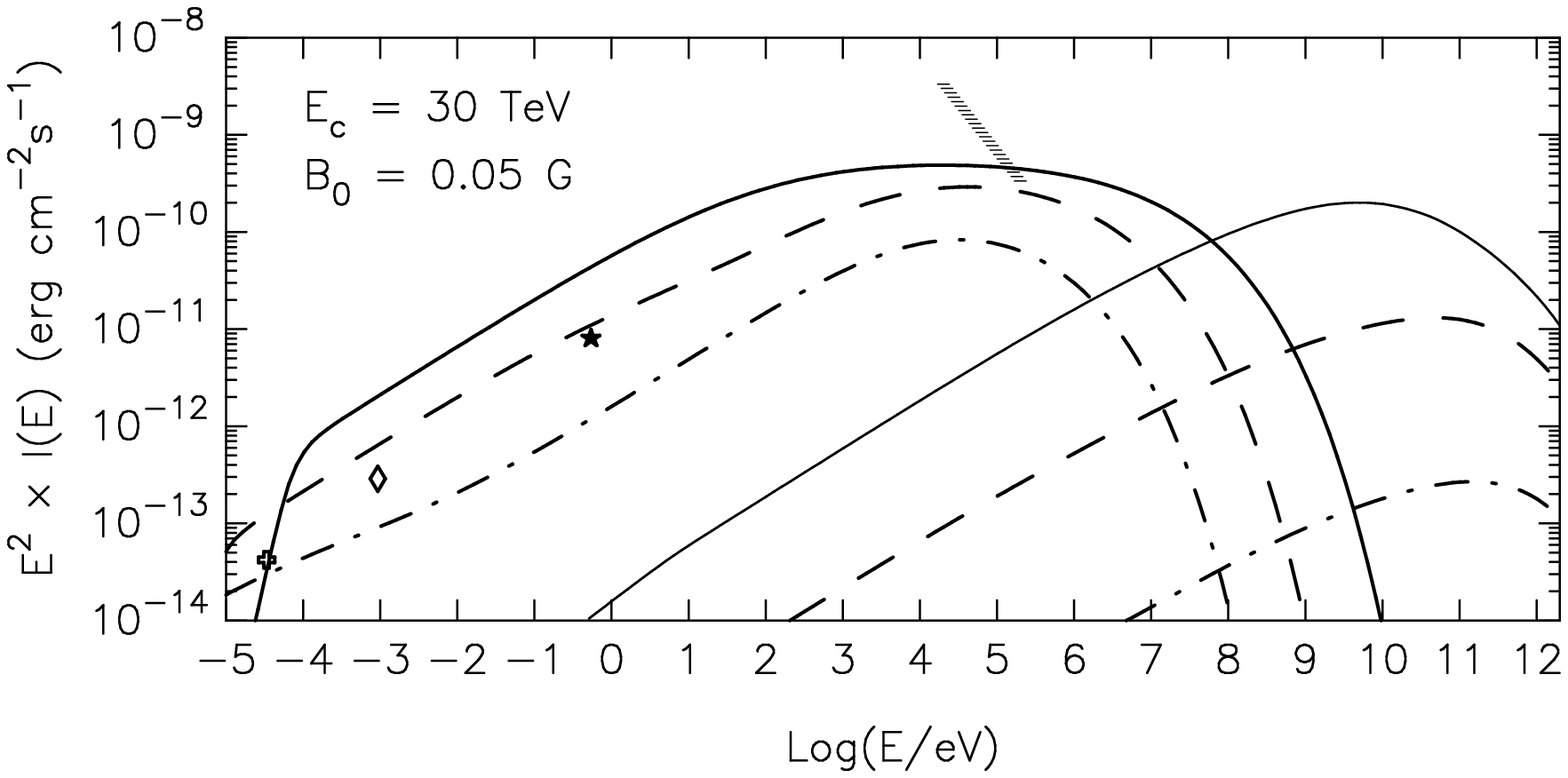}
{{\protect \bf Fig.2}:\small ~~Broad band synchrotron (heavy lines) and IC 
(thin lines) radiation fluxes 
expected  during a flare from GRS 1915+105 at $t=0.1\,\rm day$ (solid lines)
$t=1\,\rm day$ (dashed lines) and $t=10\,\rm days$ (dot dashed lines)
after relativistic ejection event. Data points correspond to the level of
maximum radio fluxes observed at $8.4\,\rm GHz$ (the cross) and $234\,\rm GHz$
(diamond) (from Rodriguez et al. 1995). The star is for the flux of IR jet 
reported by Sams et al. (1996), and the hatched region shows the level
of hard X-ray fluxes typically detected during 
the X-ray
flares of GRS~1915+105 in the energy range $\geq 20\,\rm keV$  (Harmon et al.
1997). Magnetic field $B_0 =0.05\,\rm G$ is supposed, 
which implies $L_{\rm e} = 3\times 10^{39}\,\rm erg/s$.}  
\end{figure}

\vspace{3mm}

\noindent
{\bf References}
\small

\noindent
Atoyan, A.M., and Aharonian, F.A. 1997, in preparation (AA97)

\noindent
Bodo, G., and Ghisellini, G. 1995, ApJ {\bf 441}, L69

\noindent
Castro-Tirado, A., et al. 1994, ApJS {\bf 92}, 469

\noindent
Harmon, B.A., et al. 1994, The second Compton Symposium,
AIP Conf. Proc. {\bf 304}, p.210

\noindent
Harmon, B.A., et al. 1995, Nature {\bf 374}, 703

\noindent
Harmon, B.A., et al. 1997, ApJ {\bf 477}, L85

\noindent
Hjellming, R.M., and Rupen, M.P. 1995, Nature {\bf 375}, 464

\noindent
Kardashev, N.S. 1962, Sov. Astron.-AJ {\bf 6}, 317

\noindent
Mirabel, I.F., and Rodriguez, L.F. 1994, Nature {\bf 371}, 46 (MR94)
 
\noindent
Mirabel, I.F., et al. 1997, ApJ {\bf 477}, L45 

\noindent
Rodriguez, L.F., et al. 1995, ApJS {\bf 101}, 173 

\noindent
Sams, B.J., Eckart, A., and Sunyaev, R. 1996, Nature {\bf 382}, 47

\noindent
Sazonov, S.Y., et al. 1994, Astron. Lett. {\bf 20}, 787. 

\noindent
Tingay S.J., et al. 1995, Nature {\bf 374}, 141

\noindent
Urry, M.C., and Padovani, P. 1995, PASP {\bf 107}, 803

\end{document}